\documentstyle[aps,prb,preprint]{revtex}
\tighten
\begin{document}
\title{Temperature behavior of the magnon modes of the square lattice
antiferromagnet}
\author{A.~Sherman}
\address{Institute of Physics, University of Tartu, Riia
142, 51014 Tartu, Estonia}
\author{M.~Schreiber}
\address{Institut f\"ur Physik, Technische Universit\"at,
D-09107 Chemnitz, Federal Republic of Germany}
\date{\today}
\maketitle
\begin{abstract}
A spin-wave theory of short-range order in the square lattice Heisenberg
antiferromagnet is formulated. With growing temperature from $T=0$ a
gapless mode is shown to arise simultaneously with opening a gap in the
conventional spin-wave mode. The spectral intensity is redistributed
from the latter mode to the former. For low temperatures the theory
reproduces results of the modified spin-wave theory by M.~Takahashi,
J.~E.~Hirsch {\it et al.} and without fitting parameters gives values of
observables in good agreement with Monte Carlo results in the
temperature range $0\leq T\lesssim 0.8J$ where $J$ is the exchange
constant.
\end{abstract}

\narrowtext

Properties of the spin-$\case{1}{2}$ quantum Heisenberg antiferromagnet
on a square lattice attract much attention in connection with the
investigation of cuprate-perovskite high-temperature superconductors.
In accord with the existing theories
\cite{arovas,chakra,takahashi,tang,chubukov} the spectrum of this
antiferromagnet contains the doubly degenerate (in the magnetic
Brillouin zone) magnon mode which is gapless at zero temperature. For
$T>0$ in this mode a gap opens near the center of the zone. The
appearance of this gap is connected with the short-range
antiferromagnetic ordering which is established in two-dimensional
antiferromagnets at nonzero temperature.\cite{mermin}

Of special note are the spin-wave theory of
Refs.~\onlinecite{takahashi,tang} where without fitting parameters
values of many observables were calculated in good agreement with the
exact diagonalization and Monte Carlo calculations in the temperature
range $0\leq T\lesssim 0.6J$. Results of these works can be obtained
with the mean-field decoupling of terms of the Hamiltonian which are
quartic in the magnon operators or, equivalently, by decoupling of
many-particle Green's functions in the equations of motion for the
one-particle Green's functions.\cite{sher} In the theory of
Refs.~\onlinecite{takahashi,tang} the gap in the finite-temperature
magnon spectrum appears with imposing the constraint of zero staggered
magnetization to retain the sublattice symmetry in short-range order and
to ensure zero site magnetization in the absence of magnetic field.
Notice that this spin-wave approximation is not rotationally invariant.

In the present paper we try to obtain a better approximation for the
magnon Green's functions by transferring the decoupling to higher order
equations of motion. This allows us to expand the temperature range
where the theory conforms with Monte Carlo data up to $T\approx 0.8J$.
Besides, in our theory, as soon as the temperature exceeds zero an
additional gapless mode arises simultaneously with opening a gap in the
conventional magnon mode. For large crystals and low temperatures the
spectral intensity of the gapless mode is weak and our theory reproduces
results of Refs.~\onlinecite{takahashi,tang}. With growing temperature
the spectral intensity is redistributed from the conventional to the
gapless mode.

We consider the antiferromagnetic Heisenberg model on a plane square 
lattice with the Hamiltonian
\begin{equation}
H=J\sum_{\bf la}{\bf S_l S_{l+a}}.
\label{heisenberg}\end{equation}
Here ${\bf S_l}$ is the spin-$\case{1}{2}$ operator, {\bf l} runs over
sites of one of two sublattices, and {\bf a} are vectors of the four
nearest neighbors of site zero. In the following discussion the exchange
constant $J$ is taken as the unit of energy.

The Dyson-Maleev transformation \cite{tyablikov} is used to represent
the spin operators by boson operators $a_{\bf l}$ and $b_{\bf m}$ on the
two sublattices $A$ and $B$
\begin{eqnarray}
S^-_{\bf l}&=&a^\dagger_{\bf l},\quad
  S^+_{\bf l}=\left(1-a^\dagger_{\bf l}a_{\bf l}\right)
   a_{\bf l},\quad
  S^z_{\bf l}=\frac{1}{2}-a^\dagger_{\bf l}a_{\bf l},\nonumber\\
&&\mbox{}{\bf l} \in A, \nonumber\\[-0.5ex]
&&\label{dyson} \\[-0.5ex]
S^-_{\bf m}&=&-b_{\bf m},\quad
  S^+_{\bf m}=-b^\dagger_{\bf m}
   \left(1-b^\dagger_{\bf m}b_{\bf m}\right),\nonumber\\
&&\mbox{}S^z_{\bf m}=-\frac{1}{2}+b^\dagger_{\bf m}b_{\bf m},\quad
   {\bf m} \in B. \nonumber
\end{eqnarray}
In the new notations Hamiltonian (\ref{heisenberg}) acquires the form
\begin{eqnarray}
H&=&-\frac{N}{2}+2\sum_{\bf l}a^\dagger_{\bf l}a_{\bf l}+
     2\sum_{\bf m}b^\dagger_{\bf m}b_{\bf m} \nonumber\\
 &&\mbox{}+\sum_{\bf la}\left(\frac{1}{2}a^\dagger_{\bf l}a_{\bf l}
    a_{\bf l}b_{\bf l+a}+\frac{1}{2}a^\dagger_{\bf l}
    b^\dagger_{\bf l+a}b^\dagger_{\bf l+a}b_{\bf l+a}\right.\nonumber\\
 &&\mbox{}\left.\quad\quad\quad
    -a^\dagger_{\bf l}a_{\bf l}b^\dagger_{\bf l+a}b_{\bf l+a}-
    \frac{1}{2}a^\dagger_{\bf l}b^\dagger_{\bf l+a}-
    \frac{1}{2}a_{\bf l}b_{\bf l+a}\right),
\label{hamilton}\end{eqnarray}
where $N$ is the total number of sites.

To investigate the spectrum of elementary excitations we shall calculate
the following Green's functions:
\begin{eqnarray}
D_1({\bf q}t)&=&-i\theta(t)\bigl\langle\left[a^\dagger_{\bf q}(t),
 a_{\bf q}\right]\bigr\rangle, \nonumber\\[-0.5ex]
&&\label{green}\\[-0.5ex]
D_2({\bf q}t)&=&-i\theta(t)\langle\left[b_{\bf -q}(t),a_{\bf q}
 \right]\rangle,\nonumber
\end{eqnarray}
where $a^\dagger_{\bf q}=(2/N)^{1/2}\sum_{\bf l}\exp(i{\bf ql})
a^\dagger_{\bf l}$, $b_{\bf q}=(2/N)^{1/2}$ $\sum_{\bf m}
\exp(-i{\bf qm})b_{\bf m}$ with the wave vector {\bf q} in the
magnetic Brillouin zone, $a^\dagger_{\bf q}(t)=\exp(iHt)
a^\dagger_{\bf q}\exp(-iHt)$, and angular brackets denote
thermodynamic averaging.  Equations of motion for these Green's
functions read
\begin{eqnarray}
i\frac{d}{dt}D_1({\bf q}t)&=&-\delta(t)-2D_1({\bf q}t)+
 2\gamma_{\bf q}D_2({\bf q}t)+D_3({\bf q}t), \nonumber\\[-0.3ex]
&& \label{eqmotion} \\[-0.3ex]
i\frac{d}{dt}D_2({\bf q}t)&=&2D_2({\bf q}t)-
 2\gamma_{\bf q}D_1({\bf q}t)+D_4({\bf q}t), \nonumber
\end{eqnarray}
where $\gamma_{\bf q}=\case{1}{4}\sum_{\bf a}\exp(i{\bf qa})$ and
\widetext
\begin{eqnarray*}
D_3({\bf q}t)&=&-i\theta(t)\sqrt{\frac{2}{N}}\sum_{\bf la}
 e^{i{\bf ql}}\Bigl\langle\left[\left(a^\dagger_{\bf l}(t)
 b^\dagger_{\bf l+a}(t)b_{\bf l+a}(t)-a^\dagger_{\bf l}(t)
 a_{\bf l}(t)b_{\bf l+a}(t)\right),a_{\bf q}\right]\Bigr\rangle,\\
D_4({\bf q}t)&=&-i\theta(t)\sqrt{\frac{2}{N}}\sum_{\bf ma}
 e^{i{\bf qm}}\Bigl\langle\left[\left(a^\dagger_{\bf m+a}(t)
 b^\dagger_{\bf m}(t)b_{\bf m}(t)-a^\dagger_{\bf m+a}(t)
 a_{\bf m+a}(t)b_{\bf m}(t)\right),a_{\bf q}\right]\Bigr\rangle.
\end{eqnarray*}

\narrowtext
The decoupling of the many-particle Green's functions $D_3$ and $D_4$ 
can be carried out at this stage.  If, like in Ref.~\onlinecite{tang}, 
an additional constraint of zero staggered magnetization
\begin{eqnarray}
\sum_{\bf l}&&S^z_{\bf l}-\sum_{\bf m}S^z_{\bf m}=0 \quad
 \Longrightarrow \nonumber\\
 &&\sum_{\bf l}a^\dagger_{\bf l}a_{\bf l}+
 \sum_{\bf m}b^\dagger_{\bf m}b_{\bf m}=\frac{N}{2}
\label{zsm}\end{eqnarray}
is imposed by incorporating it with a Lagrange multiplier in Hamiltonian
(\ref{hamilton}) and if the correlations $C_1=\langle a_{\bf l}b_{\bf
l+a}\rangle= \langle a^\dagger_{\bf l}b^\dagger_{\bf l+a}\rangle$ and
$\langle a^\dagger_{\bf l}a_{\bf l}\rangle=\langle b^\dagger_{\bf
m}b_{\bf m}\rangle$ are taken into account in the decoupling, results of
Refs.~\onlinecite{takahashi,tang} are reproduced.

In this paper we try to obtain a better approximation. For this purpose
we derive equations of motion for the functions $D_3$ and $D_4$ and
carry out the decoupling in the many-particle Green's functions which
appear in these equations. If in addition to the mentioned correlations
we take into account the correlations $\langle a^\dagger_{\bf l}a_{\bf
l+a_1+a_2}\rangle$ (where $\bf a_1$ and $\bf a_2$ are vectors of nearest
neighbors of site zero) and analogous correlations on the second
sublattice, we find
\widetext
\begin{eqnarray}
i\frac{d}{dt}D_3({\bf q}t)&=&4\left(C_1-\frac{1}{2}\right)\delta(t)+
 \frac{\kappa}{2}\gamma_{\bf q}D_2({\bf q}t)+\left[
 \frac{\kappa}{2}+16C_1\left(C_1-\frac{1}{2}\right)
 \left(1-\gamma^2_{\bf q}\right)\right]D_1({\bf q}t), 
 \nonumber\\[-0.5ex]
\label{dec}\\[-0.5ex]
i\frac{d}{dt}D_4({\bf q}t)&=&4\left(C_1-\frac{1}{2}\right)
 \gamma_{\bf q}\delta(t)+
 \frac{\kappa}{2}\gamma_{\bf q}D_1({\bf q}t)+\left[
 \frac{\kappa}{2}+16C_1\left(C_1-\frac{1}{2}\right)
 \left(1-\gamma^2_{\bf q}\right)\right]D_2({\bf q}t), \nonumber
\end{eqnarray}
\narrowtext\noindent
where 
$$\kappa=8\left(\sum_{\bf a_1} \bigl\langle a^\dagger_{\bf
m+a_1}a_{\bf m+a_2}\bigr\rangle \bigl\langle a_{\bf m+a_1}a^\dagger_{\bf
m+a_2}\bigr\rangle-4C_1^2\right).$$ 
Owing to the symmetry with respect to translations and rotations the
summation in $\kappa$ does not depend on {\bf m} and ${\bf a_2}$. To
derive Eqs.~(\ref{dec}) we have taken into account that in accord with
the condition of zero site magnetization in the absence of magnetic
field in short-range order
\begin{equation}
\bigl\langle a^\dagger_{\bf l}a_{\bf l}\bigr\rangle=\bigl\langle
b^\dagger_{\bf m}b_{\bf m}\bigr\rangle=\frac{1}{2}.
\label{site}\end{equation}
This condition follows also from constraint (\ref{zsm}). Substituting 
Eqs.~(\ref{dec}) into Eqs.~(\ref{eqmotion}) we get for the Fourier 
transforms of Green's functions
\widetext
\begin{eqnarray}
D_1({\bf q}\omega)&=&\frac{\left(4C_1-\omega\right)
 \left(\omega^2-\omega^2_{02}\right)-\frac{\displaystyle\kappa}
 {\displaystyle 2}\left[4\left(C_1-\frac{\displaystyle 1}{\displaystyle 
 2}\right)\left(1-\gamma^2_{\bf q}\right)-\omega\right]}
 {\left(\omega^2-\omega^2_1\right)\left(\omega^2-\omega^2_2\right)},
 \nonumber\\
&&\label{solution}\\
D_2({\bf q}\omega)&=&\frac{4\gamma_{\bf q}C_1\left(\omega^2-
 \omega^2_{02}\right)-\frac{\displaystyle\kappa}{\displaystyle 2}
 \omega\gamma_{\bf q}}
 {\left(\omega^2-\omega^2_1\right)\left(\omega^2-\omega^2_2\right)},
 \nonumber
\end{eqnarray}
where
\begin{eqnarray}
\omega^2_{1,2}&=&\frac{1}{2}\left(\omega^2_{01}+\omega^2_{02}+\kappa
 \right)\pm\sqrt{\frac{1}{4}\left(\omega^2_{01}-\omega^2_{02}
 \right)^2+\frac{\kappa}{2}\left(\omega_{01}-\omega_{02}\right)^2+
 \frac{\kappa^2}{4}\gamma^2_{\bf q}}, \nonumber\\[-1ex]
&&\label{freq}\\[-1ex]
\omega^2_{01}&=&16C_1^2\left(1-\gamma^2_{\bf q}\right),
 \quad
 \omega^2_{02}=16\left(C_1-\frac{1}{2}\right)^{\!2}\!\left(1-
 \gamma^2_{\bf q}\right).\nonumber
\end{eqnarray}

\narrowtext
Green's functions (\ref{solution}) contain two poles $\omega_1$ and
$\omega_2$ which correspond to two branches of the magnon spectrum. If
$\kappa$ is set to zero, the second pole disappears in Green's functions
and the remaining pole acquires the dispersion of the conventional linear
spin-wave theory $\omega_1=4C_1\left(1-\gamma^2_{\bf q}\right)^{1/2}$.
For $\kappa\neq 0$ a gap of the width $\kappa^{1/2}$ opens in this
branch near ${\bf q}=0$. Simultaneously the second branch acquires
finite spectral intensity which is subtracted from the intensity of the
first branch. As will be seen below, in the considered temperature range
$C_1$ is close to $\case{1}{2}$. This allows one to approximate the
dispersion of the first branch as $\omega_1\approx\left[\kappa +
16C^2_1\left(1 - \gamma^2_{\bf q}\right)\right]^{1/2}$ which, with some
change of notations, coincides with the dispersion obtained in
Refs.~\onlinecite{takahashi,tang}. Technically in those works the gap
appears in the magnon spectrum with imposing the constraints of zero
staggered or site magnetization, which are incorporated into the
Hamiltonian or free energy. In the more exact treatment of the Green's
functions the magnon gap arises without such changes of the Hamiltonian.

Parameters $C_1$ and $\kappa$ in the above equations contain
correlations which can be deduced from Green's functions
(\ref{solution}). This gives the following self-consistency conditions
for evaluating the parameters:
\begin{equation}
C_1=\frac{2}{N}\sum_{\bf q}\gamma_{\bf q}{\cal J}_{\bf q},
\label{ci}\end{equation}
\begin{equation}
\kappa=8\left(\frac{3}{4}+2K_1^2+K_2^2-4C^2_1\right),
\label{kappa}\end{equation}
where $K_1=(2/N)\sum_{\bf q}{\cal I}_{\bf q}\cos(q_x-q_y)$, $K_2=(2/N)$
$\sum_{\bf q}{\cal I}_{\bf q}\cos(2q_x)$,
\widetext
\begin{eqnarray}
{\cal J}_{\bf q}&=&\frac{4C_1\gamma_{\bf q}}{\omega^2_1-
 \omega^2_2}\left\{\frac{\omega^2_1-\omega^2_{02}}{\omega_1}\left[
 n(\omega_1)+\frac{1}{2}\right]-\frac{\omega^2_2-\omega^2_{02}}
 {\omega_2}\left[n(\omega_2)+\frac{1}{2}\right]\right\},\nonumber\\
&&\label{ij}\\
{\cal I}_{\bf q}&=&\frac{{\cal J}_{\bf q}}{\gamma_{\bf q}}-
\frac{2\kappa\left(C_1-\frac{\displaystyle 1}{\displaystyle 
 2}\right)\left(1-\gamma^2_{\bf q}\right)}{\left(\omega^2_1-
 \omega^2_2\right)}\left\{\frac{1}{\omega_1}\left[n(\omega_1)
 +\frac{1}{2}\right]-\frac{1}{\omega_2}\left[n(\omega_2)
 +\frac{1}{2}\right]\right\}, \nonumber
\end{eqnarray}
\narrowtext\noindent
$n(\omega)=\left[\exp(\omega/T)-1\right]^{-1}$, $q_x$ and $q_y$ are the
components of the wave vector {\bf q} (the intersite distance is taken
as the unit of length). In Eqs.~(\ref{ci}), (\ref{kappa}) and below
summations over wave vectors are carried out over the magnetic Brillouin
zone. Analogously, from Eqs.~(\ref{solution}) we find for the spin
correlation functions
\begin{eqnarray}
&&\bigl\langle{\bf S_lS_{l'}}\bigr\rangle=\left[\frac{2}{N}\sum_{\bf q}
 {\cal I}_{\bf q}e^{i{\bf q}({\bf l}-{\bf l'})}\right]^2-\frac{1}{4}
 \delta_{\bf ll'},\nonumber\\[-1ex]
&&\label{corr}\\[-1ex]
&&\bigl\langle{\bf S_lS_{m}}\bigr\rangle=-\left[\frac{2}{N}\sum_{\bf q}
 {\cal J}_{\bf q}e^{i{\bf q}({\bf l}-{\bf m})}\right]^2.\nonumber
\end{eqnarray}
It can be seen that correlations $\bigl\langle S^+_{\bf l}S^-_{\bf
l'(m)}\bigr\rangle$ are zero and only $\bigl\langle S^z_{\bf l} S^z_{\bf
l'(m)}\bigr\rangle$ contribute in the above expressions. Thus the
considered spin-wave approximation is not rotationally invariant.

To determine the parameters $C_1$ and $\kappa$, instead of one of
Eqs.~(\ref{ci}), (\ref{kappa}) one can use the condition of zero site
magnetization (\ref{site}) which can be rewritten as
\begin{equation}
1=\frac{2}{N}\sum_{\bf q}{\cal I}_{\bf q}.
\label{ssite}\end{equation}
Comparing results obtained with different pairs of these three equations
with Monte Carlo data, we found that the best agreement is achieved with
Eqs.~(\ref{ci}) and (\ref{ssite}). Notice also that all three equations
can be used introducing some parameter in addition to $\kappa$ and
$C_1$. In particular, such additional parameter appears if we
incorporate the constraint of zero staggered magnetization (\ref{zsm})
in Hamiltonian (\ref{hamilton}) with a Lagrange multiplier [condition
(\ref{ssite}) follows from this constraint, if one takes into account
the translation and sublattice symmetry]. This approach is a
generalization of the method of Refs.~\onlinecite{takahashi,tang} ---
results of these works can be obtained from formulas of the
three-parameter approach by setting $\kappa=0$. We postpone the
consideration of this approach to the end of the article. For now we
discuss results obtained in the two-parameter approach based on
Eqs.~(\ref{ci}) and (\ref{ssite}). Notice that the relation $\langle
{\bf S}^2_{\bf l}\rangle=\case{3}{4}$ follows from Eqs.~(\ref{corr}) and
(\ref{ssite}).

Let us first consider the case of low temperatures. As will be seen
below, $\kappa={\cal O}(N^{-2})$ for $T=0$. Since $q^2=(2\pi)^2(n^2_x+
n^2_y)/N$ where $n_x$ and $n_y$ are integers, even for the smallest
$q^2$, excluding $q^2=0$, $\omega^2_{01}$ and $\omega^2_{02}$ are much
larger than $\kappa$ for large $N$. In this case Eqs.~(\ref{ij}) can be
simplified to
\begin{eqnarray}
{\cal I}_{\bf q}&=&\frac{4C_1}{\omega_1}\left[n(\omega_1)+
 \frac{1}{2}\right]+{\cal O}(\kappa),\nonumber\\[-1ex]
&&\label{aij}\\[-1ex]
{\cal J}_{\bf q}&=&\frac{4C_1\gamma_{\bf q}}{\omega_1}\left[n(\omega_1)+
 \frac{1}{2}\right]+{\cal O}(\kappa).\nonumber
\end{eqnarray}
where $\omega_1\approx(\omega^2_{01}+\kappa)^{1/2}$ (here we took into
account that $C_1-\case{1}{2}\ll C_1$). Notice that the contribution of
the gapless magnon mode dropped out from these equations. They are also
suitable for an infinite crystal when $T\rightarrow 0$ and $\kappa$ is
exponentially small. With these ${\cal I}_{\bf q}$, ${\cal J}_{\bf q}$
and $\omega_1$ Eqs.~(\ref{ci}) and~(\ref{ssite}) come close to the
respective formulas of Ref.~\onlinecite{takahashi}. As a consequence, in
the limit $N\rightarrow\infty$, $T\rightarrow 0$ values of observables
obtained with Eqs.~(\ref{ci}), (\ref{ssite}) are similar to those found
in Ref.~\onlinecite{takahashi}. Therefore we only briefly discuss this
limit below.

Using Eqs.~(\ref{aij}), for large $N$ and $T=0$ the sum in the first 
equation (\ref{corr}) can be written as
\begin{equation}
\frac{2}{N}\sum_{\bf q}{\cal I}_{\bf q}e^{i\bf qr}=
 \frac{4C_1}{N\sqrt{\kappa}}+\frac{1}{N}\sum_{{\bf q}\neq 0}
 \frac{1}{\sqrt{1-\gamma^2_{\bf q}}}e^{i\bf qr}.
\label{sumi}\end{equation}
The sublattice magnetization $m_0=4C_1/(N\kappa^{1/2})$ is determined by
Eq.~(\ref{ssite}), $m_0=1-N^{-1}\sum_{{\bf q}\neq 0}\left(1-
\gamma^2_{\bf q}\right)^{-1/2}.$ For an infinite crystal $m_0=0.3034$
which is in good agreement with the Monte Carlo calculations.
\cite{reger,liang} As follows from the above formulas, $\kappa$ is
actually of the order of $N^{-2}$. From Eq.~(\ref{sumi}) for large {\bf
r} we get $m_0+\left(2^{1/2}\pi r\right)^{-1}$ and after analogous
transformations in the second equation (\ref{corr}) we find
\begin{equation}
\bigl\langle{\bf S_0S_{r}}\bigr\rangle\approx (-1)^r\left[m_0+
\left(\sqrt{2}\pi r\right)^{-1}\right]^2,
\label{asymp}\end{equation}
where $(-1)^r=+1$ or $-1$ depending on whether the sites {\bf 0} and
{\bf r} belong to the same or different sublattices. In Table~\ref{tabi}
the zero-temperature spin correlations $C_{\bf r}=\langle{\bf
S_0S_r}\rangle$, obtained by numerical solution of Eqs.~(\ref{ci})
and~(\ref{ssite}) (see below), are compared with results of the
projected Monte Carlo method. \cite{liang} The values agree nicely.

For $N\rightarrow\infty$, $T\rightarrow 0$ we find from Eqs.~(\ref{ci}),
(\ref{ssite}) and (\ref{aij})
\begin{eqnarray}
&&C_1=m_1-\frac{4}{\pi}\zeta(3)\left(\frac{T}{4m_1}\right)^3+{\cal
 O}(T^5),\nonumber\\[-0.5ex]
&&\label{lowt}\\[-0.5ex]
&&\sqrt{\kappa}=T\exp\left(-\frac{2\pi m_0m_1}{T}\right)
 \left[1+{\cal O}(T^2)\right],\nonumber
\end{eqnarray}
where $m_1=1-N^{-1}\sum_{\bf q}\left(1-\gamma^2_{\bf q}\right)^{1/2} =
0.57897$ and $\zeta(x)$ is the Riemann zeta function. For large {\bf r}
the sums in the spin correlations $\langle{\bf S_0S_r}\rangle$ 
(\ref{corr}) can be rewritten as
$$\frac{T}{(2\pi)^2C_1}\int\!\!\!\!\int\!\frac{d^2q}{q^2+(2\xi)^{-2}}
 e^{i\bf qr}
 \approx\frac{T}{2C_1}\sqrt{\frac{\xi}{\pi r}}\exp\left(-\frac{r}{2\xi}
 \right)$$
with the correlation length
\begin{equation}
\xi=C_1\sqrt{\frac{2}{\kappa}}=\frac{\sqrt{2}m_1}{T}\exp\left(
 \frac{2\pi m_0m_1}{T}\right)\left[1+{\cal O}(T^2)\right].
\label{cl}\end{equation}
This value of the correlation length is in agreement with results
obtained in Refs.~\onlinecite{arovas,chakra}.

To solve Eqs.~(\ref{ci}) and~(\ref{ssite}) for arbitrary $T$ and $N$ we
determined the minimum of the function
$$F(C_1,\kappa)=\left(1-\frac{2}{N}\sum_{\bf q}{\cal I}_{\bf q}
 \right)^{\!2}\!
 +\left(1-\frac{2}{NC_1}\sum_{\bf q}\gamma_{\bf q}{\cal J}_{\bf q}
 \right)^{\!2}\!,$$
which is constructed from squares of the differences of the right and
left sides of these equations. The iteration procedure with the steepest
descent method was continued until $F$ was less than 10$^{-12}$. As an
example, parameters obtained by this procedure for a 20$\times$20
lattice are given in Table~\ref{tabii}. These parameters were used for
calculating the static uniform susceptibility
\begin{eqnarray}
\chi&=&\frac{1}{T}\sum_{\bf r}\bigl\langle S^z_{\bf 0}S^z_{\bf r}
 \bigr\rangle=\frac{1}{3T}\sum_{\bf r}\bigl\langle{\bf S_0S_r}
 \bigr\rangle \nonumber\\
 &=&\frac{1}{3T}\left[\frac{2}{N}\sum_{\bf q}\left({\cal
 I}^2_{\bf q}-{\cal J}^2_{\bf q}\right)-\frac{1}{4}\right]
\label{chi}\end{eqnarray}
and the energy per spin
\begin{equation}
E=2\bigl\langle{\bf S_lS_{l+a}}\bigr\rangle=-2C_1^2.
\label{energy}\end{equation}
Results for a 20$\times$20 lattice are shown in Figs.~\ref{figi} and
\ref{figii} together with the data obtained in the Monte Carlo
calculations \cite{okabe,makivic} and in the modified spin-wave theory
of Refs.~\onlinecite{takahashi,tang}. As seen from the figures, results
obtained in the spin-wave approximations of the present work and of
Refs.~\onlinecite{takahashi,tang} are close and are in good agreement
with the Monte Carlo results in the considered temperature range. The
size dependence of $\chi$ and $E$ calculated with the above formulas is
negligible for large enough lattices --- the difference between values
obtained for a 40$\times$40 lattice and those shown in Figs.~\ref{figi}
and~\ref{figii} is less than the size of symbols in these figures.
Notice also that zero frequency of the second branch at ${\bf q}=0$ does
not lead to divergencies in the above formulas as the respective
numerators in Eqs.~(\ref{ij}) approach zero sufficiently rapidly when
${\bf q} \rightarrow 0$. 

As seen from Table~\ref{tabii}, starting from low temperatures the gap
determined by $\kappa$ grows with $T$. With this growth the spectral
intensity of the second branch and its contribution to the observables
increases. However, near $T=0.5$ $\kappa$ starts to decrease. For $T
\approx 0.7$, when $\kappa\rightarrow 0$ and $C_1 \approx 0.5$, the
considered two-parameter approximation ceases to work --- it is
impossible to find parameters for which Eqs.~(\ref{ci}) and
(\ref{ssite}) are fulfilled with high accuracy.

Now let us consider the three-parameter approximation mentioned above.
The additional parameter is the Lagrange multiplier $\lambda$ with which
the terms $\sum_{\bf l}a^\dagger_{\bf l}a_{\bf l}+\sum_{\bf m}
b^\dagger_{\bf m}b_{\bf m}$ are added to Hamiltonian (\ref{hamilton}) to
ensure the fulfillment of constraint~(\ref{zsm}). With these additional
terms $D_1({\bf q}\omega)$ in Eq.~(\ref{solution}), $\kappa$ in
Eq.~(\ref{kappa}) and ${\cal I}_{\bf q}$ in Eq.~(\ref{ij}) are slightly
changed

\widetext
\[\hfill
D_1({\bf q}\omega)=\frac{\left(4C_1\eta^{-1}-\omega\right)
 \left(\omega^2-\omega^2_{02}\right)-\frac{\displaystyle\kappa}
 {\displaystyle 2}\left[4\left(C_1-\frac{\displaystyle 1}{\displaystyle
 2}\right)\left(1-\gamma^2_{\bf q}\right)-\omega\right]}
 {\left(\omega^2-\omega^2_1\right)\left(\omega^2-\omega^2_2\right)},
\hfill (\ref{solution}')\]
\[\hfill
\kappa=8\left[\frac{3}{4}+2K_1^2+K_2^2+4C^2_1\left(1-\frac{2}{\eta}
 \right)\right],
\hfill (\ref{kappa}')\]
\vspace*{-2ex}\[\hfill
{\cal I}_{\bf q}=\frac{{\cal J}_{\bf q}}{\eta\gamma_{\bf q}}-
\frac{2\kappa\left(C_1-\frac{\displaystyle 1}{\displaystyle 
 2}\right)\left(1-\gamma^2_{\bf q}\right)}{\left(\omega^2_1-
 \omega^2_2\right)}\left\{\frac{1}{\omega_1}\left[n(\omega_1)
 +\frac{1}{2}\right]-\frac{1}{\omega_2}\left[n(\omega_2)
 +\frac{1}{2}\right]\right\},\hfill (\ref{ij}')\]
whereas $D_2({\bf q}\omega)$ in Eq.~(\ref{solution}), ${\cal J}_{\bf q}$
in Eq.~(\ref{ij}), Eqs.~(\ref{ci}), (\ref{ssite}), (\ref{chi}) and
(\ref{energy}) retain their form. In these equations $\eta=[1-\lambda/
(4C_1)]^{-1}$,
\[\hfill
\begin{array}{ll}
&\omega^2_{1,2}=\frac{\displaystyle 1}{\displaystyle 2}\left(
 \omega^2_{01}+\omega^2_{02}+\kappa\right)\\[2ex]
&\quad\quad \pm\sqrt{\frac{\displaystyle 1}{\displaystyle 4}
 \left(\omega^2_{01}-\omega^2_{02}\right)^2+
 \frac{\displaystyle\kappa}{\displaystyle 2}
 \left(\omega^2_{01}+\omega^2_{02}\right)-16C_1\left(C_1-
 \frac{\displaystyle 1}{\displaystyle 2}\right)
 \frac{\displaystyle\kappa}{\displaystyle\eta}
 \left(1-\gamma^2_{\bf q}\right)+\frac{\displaystyle\kappa^2}
 {\displaystyle 4}\gamma^2_{\bf q}},\\[2ex]
&\omega^2_{01}=16C_1^2\left(\frac{\displaystyle 1}{\displaystyle\eta^2}
 -\gamma^2_{\bf q}\right), \quad 
 \omega^2_{02}=16\left(C_1-\frac{\displaystyle 1}{\displaystyle 2}
 \right)^2\left(1-\gamma^2_{\bf q}\right).
\end{array} \hfill (\ref{freq}')
\]

\narrowtext
To solve Eqs.~(\ref{ci}), (\ref{kappa}') and (\ref{ssite}) we determined
the minimum of the function $F'(C_1,\kappa,\eta)$, constructed from
squares of the differences of left and right sides of these equations,
by the steepest descent method. The static uniform susceptibility and
the energy per spin calculated with the parameters obtained in this way
for a 20$\times$20 lattice are shown in Figs.~\ref{figi}
and~\ref{figii}. As seen from the figures, in comparison with the
two-parameter approximation the three-parameter approach agrees slightly
better with the Monte Carlo results and is applicable in the wider
temperature range $0\leq T\lesssim 0.8$. However, the convergence of the
steepest descent method in the three-parameter approximation is much
worse than in the two-parameter one. We connect this with the
observation that both approaches give similar pictures of magnon modes
and in the three-parameter approximation the two parameters $\kappa$ and
$\lambda$ (or $\eta$) determine one physical quantity --- the magnon
gap. In this approximation the worsened convergence is connected with
the flat minimum of $F'$ considered as the function of $\kappa$ and
$\lambda$.

In the present work we have found the gapless mode in the state with
short-range order. In ordered states excitations of such type correspond
to Goldstone's mode and point to the existence of the continuous
degeneracy of these states.\cite{forster} The observation of the gapless
mode in the considered disordered state may also be connected with the
continuous degeneracy of this state. In some respects a similar magnon
spectrum of short-range order was obtained in
Refs.~\onlinecite{shimahara,sokol}. When considered in the magnetic
Brillouin zone the spectrum consists also of two modes one of which has
a gap near ${\bf q}=0$ and another is gapless. However, the shape of
these branches, their relative spectral intensities and temperature
behavior differ essentially from those obtained here. The spin-wave
approximations of Refs.~\onlinecite{shimahara,sokol} are rotationally
invariant but are not self-consistent --- correction coefficients and
data of other calculations are needed to fit results to experiment in
the temperature range $T\lesssim 1$.

In conclusion, the spin-wave theory of short-range order in the square
lattice Heisenberg antiferromagnet was formulated. In agreement with
previously obtained results we found that as soon as the temperature
exceeds zero and long-range order gives way to short-range order a gap
opens in the conventional magnon mode near ${\bf q}=0$. We found
additionally that a new gapless mode arises simultaneously with opening
the gap. With growing temperature the spectral intensity is
redistributed from the conventional to the gapless mode. We considered
spin-wave approximations with two and three self-consistently determined
parameters. For low temperatures and large crystals the theory
reproduces results of the modified spin-wave theory of
Refs.~\onlinecite{takahashi,tang}. Without fitting parameters our
calculations give values of the static uniform susceptibility and energy
per spin in good agreement with Monte Carlo results in the temperature
range $0\leq T\lesssim 0.8J$ where $J$ is the exchange constant.

\acknowledgements
This work was partially supported by the ESF grant No.~2688 and by the
WTZ grant (Project EST-003-98) of the BMBF.

\begin{figure}\caption{The static uniform susceptibility obtained in the
Monte Carlo simulation for a 12$\times$12 lattice \protect\cite{okabe}
($\bullet$), in the modified spin-wave approximation of
Refs.~\protect\onlinecite{takahashi,tang} ($\circ$) and in the
two-parameter ($+$) and three-parameter ($\times$) spin-wave
approximations of the present work. In the spin-wave calculations a
20$\times$20 lattice was used.}\label{figi}\end{figure}

\begin{figure}\caption{The energy per spin obtained in the Monte Carlo
simulation \protect\cite{makivic} ($\bullet$), in the modified spin-wave
approximations of Refs.~\protect\onlinecite{takahashi,tang} ($\circ$)
and in the two-parameter ($+$) and three-parameter ($\times$) spin-wave
approximations of the present work. In the spin-wave calculations a
20$\times$20 lattice was used.} \label{figii}\end{figure}

\begin{table}
\caption{The zero-temperature spin correlations $C_{\bf r}$ obtained 
with the two-parameter spin-wave approximation (SW) for a 20$\times$20
lattice in comparison with the projected Monte Carlo data (PMC).
\protect\cite{liang}
}
\label{tabi}
\begin{tabular}{|c|c|c|}
           &\hphantom{aa}PMC\hphantom{aa}
           &\hphantom{aa}SW\hphantom{aa}\\ \hline
\hphantom{aa}$C_{1,0}$\hphantom{aa}&-0.3348&-0.3354 \\
$C_{1,1}$  &  0.2028  &  0.2016 \\
$C_{2,0}$  &  0.1772  &  0.1751 \\
$C_{2,1}$  & -0.1671  & -0.1648 \\
$C_{2,2}$  &  0.1475  &  0.1454 \\
$C_{3,0}$  & -0.1491  & -0.1461 \\
$C_{3,1}$  &  0.1430  &  0.1404 \\
\end{tabular}
\end{table}

\begin{table}
\caption{Parameters $C_1$ and $\kappa$ obtained from
Eqs.~(\protect\ref{ci}) and~(\protect\ref{ssite}) for a 20$\times$20
lattice.
}
\label{tabii}
\begin{tabular}{|c|c|c|}
$T$       &     $C_1$     &     $\kappa$ \\ \hline
10$^{-5}$ &   $0.57912$ &   $2.773\cdot 10^{-4}$ \\
$0.1$     &   $0.57904$ &   $4.685\cdot 10^{-4}$ \\
$0.2$     &   $0.57826$ &   $9.390\cdot 10^{-4}$ \\
$0.3$     &   $0.57543$ &   $1.443\cdot 10^{-3}$ \\
$0.4$     &   $0.56819$ &   $1.941\cdot 10^{-3}$ \\
$0.5$     &   $0.55409$ &   $2.303\cdot 10^{-3}$ \\
$0.6$     &   $0.53037$ &   $2.093\cdot 10^{-3}$ \\
$0.7$     &   $0.50003$ &   $3.664\cdot 10^{-6}$ \\
\end{tabular}
\end{table}
\end{document}